\documentclass[twocolumn,aps,showpacs,prb,tightenlines,amsmath,amssymb]{revtex4}
\usepackage{bm}
\usepackage{graphicx}              
\usepackage{amssymb}               
\usepackage{amsmath}                     
\usepackage{colordvi}
\usepackage{calrsfs} \newcommand{\bgreek}[1]{\mbox{\boldmath$#1$\unboldmath}}
\begin{document}   

\title{Valley depolarization due to inter- and intra-valley electron-hole exchange interactions in monolayer MoS$_{2}$}
\author{T. Yu}
\author{M. W. Wu}
\thanks{Author to whom correspondence should be addressed}
\email{mwwu@ustc.edu.cn.}
\affiliation{Hefei National Laboratory for Physical Sciences at
  Microscale and Department of Physics, 
University of Science and Technology of China, Hefei,
  Anhui, 230026, China} 
\date{\today}

\begin{abstract} 
We investigate the valley depolarization due to the electron-hole exchange
interaction in monolayer
MoS$_{2}$. Both the long- and short-range parts of the intra- and
inter-valley electron-hole exchange interactions are calculated. We find
that both the long- and short-range exchange interactions can 
cause the inter- and intra-valley bright exciton transitions. With
the intra-valley bright exciton transition channel nearly 
forbidden due to the large splitting of the valence
bands, the inter-valley channel due to the 
exchange interaction can cause the valley depolarization efficiently by
the Maialle-Silva-Sham
    mechanism [Phys. Rev. B {\bf 47}, 15776 (1993)]. With
only the long-range exchange interaction, the calculations show good agreement with the recent valley polarization
experiments, including the time-resolved valley
 polarization measurement, the
pump-probe experiment and the steady-state
 PL polarization measurement. We further show that for the 
A-exciton with large (small) center-of-mass momentum, the long-range exchange 
interaction can cause the {\em fast} ({\em slow}) inter-valley exciton
transition.
\end{abstract}
%71.35.-y Excitons
%71.70.Gm Exchange interactions, energy-level splitting
%78.67.-n Optical properties of low-dimensional,
% mesoscopic, and nanoscale materials and structures
%71.10.-w theories and models of surfaces, interfaces, and thin films

\maketitle 

\section{Introduction} 
\begin{figure}[htb]
  {\includegraphics[width=8cm]{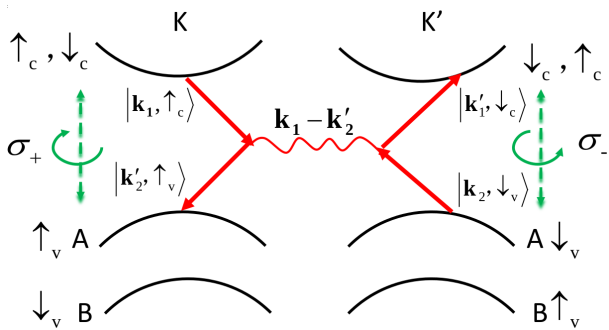}}
  \caption{(Color online) Schematic diagram of the bright exciton transition
    channel between
    the K and K$'$ valleys. The Feynman diagram shows that 
due to the short-range exchange
    interaction, the electrons in the states $|{\bf k}_1,
 \uparrow_c, {\rm K}\rangle$ and $|{\bf k}_2, \downarrow_v, {\rm K'}\rangle$ are
 scattered to the states $|{\bf k}_2', \uparrow_v, {\rm K}\rangle$ and $|{\bf
   k}_1', \downarrow_c, {\rm K'}\rangle$, respectively. 
Consequently, the inter-valley bright exciton
 transition is realized by the virtual
 recombination of a bright exciton in the K
 valley and then generation in the K$'$ valley or vice versa.}
% Note that $\uparrow_v$
% ($\downarrow_v$) denotes the electron spin in the valence band, corresponding
% the hole spin $\downarrow_h$ ($\uparrow_h$) in the following.}
\label{figyw1}
\end{figure}

Monolayer MoS$_2$ has attracted intense interest due to
its remarkable electrical and optical properties  from its unique energy band
structure (shown in Fig.~\ref{figyw1}) very 
recently.\cite{ele_experiment,ele_theory,electric_3,valley_wang6,
valley_wang18,valley_wang20,valley_wang24,direct_gap1,directgap_wang11,
direct_gap3,directgap_wang23,direct_gap5,direct_gap6,
splitting_wang27,splitting_wang31,splitting_wang37,splitting_wang38}
Unlike its bulk form, monolayer MoS$_2$ has
direct gaps at the inequivalent K and K$'$ points of 
the hexagonal Brillouin Zone,\cite{direct_gap1,directgap_wang11,
direct_gap3,directgap_wang23,direct_gap5,direct_gap6} which have been
confirmed in the photoluminescence (PL) experiments.\cite{valley_wang6,
valley_wang18,valley_wang20,valley_wang24}
 In addition, due to the
space inversion asymmetry and the strong spin-orbit coupling originated
from the {\em d}-orbitals of the heavy metal atoms, the valence
bands are splitted by about 160
meV.\cite{directgap_wang11,valley_wang20,directgap_wang23,
splitting_wang27,splitting_wang31,splitting_wang37,splitting_wang38}
Therefore, it shows two excitonic transitions A ($\approx1.9$  eV) and B ($\approx2.1$ eV) from the K or K$'$
point in the light absorption.\cite{absorption_Mak,valley_wang24,
absorption_Kioseoglou,valley_wang18,absorption_wang15,absorption_Marie} 
Moreover, the chiral optical valley
selection rule in this system leads to the selective excitation of carriers in only one of
these valleys depending on the helicity of circularly polarized light, with $\sigma_{+}$ or $\sigma_{-}$
light being directly associated with the K or K$'$ valley.\cite{valley_wang6,
valley_wang18,valley_wang20}
Accordingly, the spin polarization can be realized due to the splitting of the
valence bands.\cite{valley_wang6,
valley_wang18,valley_wang20,valley_wang24} Therefore, monolayer MoS$_2$ provides
 an ideal platform to study the semiconductor valley
physics (valleytronics). 

It has been theoretically predicted that high valley polarization 
up to $100\%$ can be realized in monolayer 
MoS$_2$.\cite{direct_gap1,valley_wang20,absorption_Mak,valley_wang24,
absorption_Kioseoglou,valley_wang18,absorption_wang15,absorption_Marie}
However, recent valley polarization experiments in
monolayer MoS$_2$ with A-exciton pumped, including the time-resolved valley
 polarization measurement,\cite{absorption_Marie} the
pump-probe experiment\cite{many_body,CD} and the steady-state
 PL polarization measurement,\cite{direct_gap1,valley_wang20,absorption_Mak,valley_wang24,
absorption_Kioseoglou,valley_wang18,absorption_wang15}
 suggest that there exists fast valley depolarization. For the time-resolved valley
 polarization measurement, the observation of the excitonic signal in the K$'$
 valley is immediate after the A-exciton pumped in the K valley and a finite
 valley polarization (about $50\%$ at 4 K) is measured during the
 A-exciton lifetime.\cite{absorption_Marie} For the pump-probe experiment, it shows that there also
 exists fast inter-valley exciton transition and finite residue valley
 polarization which lasts for tens of picoseconds.\cite{many_body,CD} In the steady-state measurements of the PL polarization,
 a wide range of valley polarizations from $30\%$ to $100\%$ are reported with the
resonantly pumping energy $E\approx 1.96$ eV for the A exciton at low temperature.\cite{absorption_Mak,valley_wang24,
absorption_Kioseoglou,valley_wang18,absorption_wang15,absorption_Marie} 
It was claimed that the valley depolarization originates from the
electron/hole spin relaxation due to the D'yakonov-Perel' (DP)\cite{DP}
and Elliott-Yafet (EY)\cite{Yafet,Elliott} mechanisms.\cite{absorption_Mak,valley_wang24,absorption_Kioseoglou,
valley_wang18,absorption_wang15,absorption_Marie} 
Therefore, after the inter-valley
 scattering including the
 electron-phonon\cite{absorption_Mak,absorption_Kioseoglou,valley_wang18,absorption_Marie}
 and/or short-range
impurity scatterings, the spin relaxation of the electron and hole
 can cause the bright exciton transition between the K and
K$'$ valleys and hence the PL depolarization.
  However, for the DP mechanism, in the {\em intrinsic} situation, it
cannot cause any spin relaxation because the out-of-plane component of the
electron or hole spin is conserved;\cite{valley_wang20,wang_relaxation1,splitting_wang39,tightbinding}
 in the {\em extrinsic} situation, the flexural deformations can
cause the spin relaxation of carriers but the spin relaxation time
is in the order
of nanoseconds.\cite{referee,flexural} For the EY mechanism, the
  out-of-plane components of the electron
  or hole spin are conserved in the intrinsic situation and only the
  extrinsic influences can cause the spin
  relaxation.\cite{valley_wang20,wang_relaxation1,splitting_wang39,tightbinding,referee}
 It is calculated
 that the spin relaxation time of the
  out-of-plane component is also in the order of nanoseconds at low
temperature with low impurity density.\cite{wang_relaxation2}
Accordingly, the exciton transition time due to the DP and EY mechanisms
 is much longer than its lifetime, which
 is in the order of picoseconds,\cite{absorption_Marie,CD14} and
 hence the DP and EY mechanisms cannot cause the PL depolarization
 effectively.

 In this paper, we show that the electron-hole (e-h) exchange interaction
 can cause the valley depolarization efficiently due to the Maialle-Silva-Sham
   (MSS) mechanism\cite{Sham1,Sham2} based on the kinetic spin Bloch equations
   (KSBEs).\cite{broadening1,broadening2}
 We show that both
   the long-range (L-R) and short-rang (S-R) parts of the exchange interactions
   can cause the inter- and intra-valley bright exciton transitions. 
 However, the intra-valley bright exciton transition channel
is nearly forbidden due to the large splitting of the valence
bands and only the inter-valley exchange interaction can cause the valley
depolarization efficiently. This inter-valley bright exciton transition process is schematically shown in
 Fig.~{\ref{figyw1}}, in which electrons in the conduction band of the K valley
 and valence band in the K$'$ valley are
 scattered to the valence band in the K valley and conduction band in the K$'$
 valley, respectively. This process can also be treated as the result of virtual
 recombination of a bright exciton in the K
 valley and generation in the K$'$ valley, or vice versa. We further show that for the 
A-exciton with large center-of-mass momentum, the L-R exchange 
interaction can cause the {\em fast} inter-valley exciton transition.
 This explains the fast emergence of the excitonic signal in the K$'$
valley with the A-exciton pumped in K valley in the experiments.\cite{absorption_Marie,many_body,CD}
However, for the 
A-exciton with small center-of-mass momentum, the inter-valley exciton
transition is relatively {\em slow}, which leads to the existence of the residue valley
polarization which lasting for tens of picoseconds in the experiments.\cite{many_body,CD}

This paper is organized as follows. In Sec.~{\ref{model}}, we set up the
  model and lay out the formalism. In Sec.~{\ref{Model_A}}, we derive the L-R
  and S-R parts
 of the exciton exchange interaction Hamiltonian and show
  that both the L-R and S-R exchange interactions can cause the inter-valley
  exciton transition. In Sec.~{\ref{Model_B}}, we present the KSBEs and
    compare the theoretical results with the experimental ones.
 We conclude and discuss in Sec.~{\ref{summary}}.

\section{Model and Formalism} 
\label{model}
\subsection{E-h exchange interaction}
\label{Model_A}
In this work, we study the e-h exchange interaction for the direct excitons
 in monolayer MoS$_2$ based on the lowest four band ${\bf
   k}\cdot {\bf p}$ Hamiltonian for the K (K$'$) 
valley:\cite{valley_wang20,splitting_wang39,tightbinding}
\begin{equation}
\label{hamiltonian}
\hat{H}=at(\tau
k_x\hat{\sigma}_x+k_y\hat{\sigma}_y)+\Delta\hat{\sigma}_z/2-\lambda
\tau(\hat{\sigma}_z-1)\hat{s}_z/2.
\end{equation}
Here, $a$ is the lattice constant, $t$ represents
the effective hopping integral; $\tau$ is the valley index for K ($\tau=1$) and
K$'$ ($\tau=-1$) valleys;  $\hat{\sigma}$ stand for the Pauli
matrices for the two basis functions [$c$ ($v$) indicates conduction (valence) band]
\begin{equation}
|\phi_c\rangle=|d_{z^2}\rangle,~~~~|\phi_v^{\tau}\rangle=\frac{1}{\sqrt{2}}\big(|d_{x^2-y^2}\rangle+i\tau|d_{xy}\rangle\big);
\label{basis}
\end{equation}
  $\Delta$ is the energy gap; $2\lambda$ denotes the
spin splitting of the valence bands and $\hat{s}_z$ is the
Pauli matrix for spin.\cite{valley_wang20} Using above Hamiltonian
  Eq.~(\ref{hamiltonian}),
 the effective Hamiltonian for
  the exciton envelop function in the
  coordinate space are
derived following the previous works within the framework of the effective-mass
approximation [refer to Eq.~(\ref{effective-mass})],\cite{exchange,Sham1,dot} shown in Appendix~\ref{AA}.
Based on this effective Hamiltonian Eq.~(\ref{effective_H}), by treating MoS$_2$ as ideal
two-dimensional (2D) material,\cite{2D} the basis functions for the
direct excitons can be expressed in terms of the
exciton ground state
\begin{equation}
|\mu_e,\nu_h,{\bf P}, {\rm {\bf K}}\rangle=\Psi_{\rm {\bf K}}^{\mu_e}({\bf
  r}_e)\tilde{\Psi}_{\rm {\bf K}}^{\nu_h}({\bf r}_h)\phi^{\rm 2D}_{1s}({\bgreek \rho})e^{i{\bf
      P}\cdot{\bf R}}/\sqrt{S}
\label{K_valley}
\end{equation} 
for the K valley and 
\begin{equation}
|\mu_e',\nu_h',{\bf P'}, {\rm {\bf K}'}\rangle=\Psi_{\rm {\bf K}'}^{\mu_e'}({\bf
  r}_e)\tilde{\Psi}_{\rm {\bf K}'}^{\nu_h'}({\bf r}_h)\phi^{\rm 2D}_{1s}({\bgreek \rho})e^{i{\bf
      P'}\cdot{\bf R}}/\sqrt{S}
\label{K'_valley}
\end{equation}
for the K$'$ valley. In Eqs.~({\ref {K_valley}}) and ({\ref {K'_valley}}),
$\Psi_{\rm {\bf K}(\rm {\bf K}')}^{\mu_e(\mu_e')}({\bf r}_e)$
and $\tilde{\Psi}_{\rm {\bf K}(\rm {\bf K}')}^{\nu_h(\nu_h')}({\bf r}_h)$ are the electron and hole
Bloch wave functions with $\mu_e$ $(\mu_e')$ and $\nu_h$ $(\nu_h')$ denoting the
electron spin
 in the conduction bands and
hole spin in the valence bands (note $\nu_h=-\nu_e$ 
with $\nu_e$ being the electron
spin in the valence bands),
respectively; $\phi^{\rm 2D}_{1s}({\bgreek \rho})=\sqrt{8/{\pi a_B^2}}\exp({-2\rho/a_B})$ is the 2D hydrogenic exciton
ground-state wave
function with ${\bgreek \rho}={\bf r}_e-{\bf r}_h$ standing for the relative coordinate of
the electron and hole, and $a_B$ is the exciton radius; the last factor represents the center-of-mass motion of
the exciton with $S$, ${\bf P}$ $({\bf P'})$ and ${\bf R}$ representing the area of the 2D plane of
MoS$_2$, the center-of-mass wavevector of the electron-hole pair and its
center-of-mass position, respectively. 

The exchange interaction is divided into the L-R and S-R
parts. Their matrix elements between two exciton ground states are further
  derived based on the exciton Hamiltonian in Appendix~\ref{AA} by expanding 
  the exciton envelop function using the exciton ground states,\cite{exchange,Sham1,dot} which are shown as
follows.

\subsubsection{L-R part}

For the L-R part of the exchange interaction, there exist
  matrix elements for the excitons in a single valley or between K and K$'$
  valleys.\cite{exchange,Sham1} The matrix
elements between two exciton ground states $|m,n,{\bf P}, {\bf k}_0\rangle$
and $|m',n',{\bf P'}, {\bf k}_0'\rangle$ are expressed as 
\begin{equation}
B^{\rm L\mbox{-}R}_{{\bf k}_0\mbox{-}{\bf k}_0'}=\frac{e^2}{2\kappa
    \varepsilon_0|{\bf P}|}\delta_{{\bf P},{\bf
      P}'}|\phi_{1s}^{\rm
    2D}(0)|^2Q^{{\bf k}_0\mbox{-}{\bf k}_0'}_{m'\Theta n
    \atop \Theta n' m}({\bf P}),
\label{long_range}
\end{equation}
with $\varepsilon_0$ and $\kappa$ standing for the vacuum permittivity and relative
dielectric constant,
respectively; ${\bf k}_0({\bf k}_0')$ representing ${\bf K}$ or ${\bf K}'$;
\begin{eqnarray}
\nonumber 
&&Q^{{\bf k}_0\mbox{-}{\bf k}_0'}_{m'\Theta n
    \atop \Theta n' m}({\bf
    P})=\frac{\hbar^2}{m_0^2}\big[{\bf
  P}\cdot{\bgreek \pi_{m'\Theta n'}({\bf k}_0)}\big]\big[{\bf
  P}'\cdot{\bgreek \pi_{\Theta n m}({\bf k}_0')}\big]\\
&&\mbox{}\times\frac{1}{\big[E_m({\bf k}_0)-E_n({\bf k}_0)\big]\big[E_{m'}({\bf
    k}_0')-E_{n'}({\bf k}_0')\big]}.
\end{eqnarray}
Here, $m_0$ is the free electron mass; ${\bgreek \pi_{m' \Theta
    n'}({\bf k}_0)}$ and ${\bgreek \pi_{\Theta n m}({\bf k}_0')}$ come from the ${\bf k}\cdot{\bf p}$ matrix
elements in the Hamiltonian [Eq.~(\ref{hamiltonian})] 
with $\Theta$ being the time reversal operator (refer to Appendix~\ref{AA});
 $E_{\mu}({\bf k})$ and $E_{\nu}({\bf k}')$ 
are the edge energies of the conduction band with electron spin $\mu$ and
valence band with hole spin $\nu$.

For the intra-valley exciton exchange interaction, according to Eq.~(\ref{long_range}), with the spin bases $(\mu_e,\nu_h)$
in the order $(\uparrow_e,\downarrow_h), (\downarrow_e,\downarrow_h), (\uparrow_e,\uparrow_h),
(\downarrow_e,\uparrow_h)$,
the L-R part of the exchange
  interaction between two exciton states $|m,n,{\bf P}, {\bf k}_0\rangle$
and $|m',n',{\bf P'}, {\bf k}_0\rangle$ is written as 
\begin{equation}
H^{\rm L\mbox{-}R}_{{\bf k}_0\mbox{-}{\bf k}_0}=C\delta_{\bf P, P'}|{\bf P}|\left(\begin{array}{cccc}
\alpha(\tau) & 0 & 0 & \beta\\
0 & 0 & 0  & 0\\
0 & 0  & 0  & 0\\
\beta & 0 & 0 & \alpha(-\tau)
\end{array}\right).
\label{longrange_intra}
\end{equation}
Here, $C=e^2/(2\kappa
    \varepsilon_0)|\phi_{1s}^{\rm 2D}(0)|^2$ and $\tau=1(-1)$ for ${\bf
    k}_0={\bf K}({\bf K}')$. The parameters
$\alpha(\tau)=a^2t^2/(\Delta-\tau\lambda)^2$ and $\beta=a^2t^2/(\Delta^2-\lambda^2)$    
for MoS$_2$,
WS$_2$, MoSe$_2$ and WSe$_2$ are shown in Table~\ref{parameter} calculated with
the material parameters taken from Ref.~{\onlinecite{valley_wang20}}.

\begin{table}[htb]
  \caption{Material parameters $\alpha(\tau)$ and $\beta$ for MoS$_2$,
WS$_2$, MoSe$_2$ and WSe$_2$ with the unit being ${\rm \AA}^2$.}
  \label{parameter} 
  \begin{tabular}{l l l l}
    \hline
    \hline
    &\;\;\;\;{$\alpha(1)$}\;\;\;&\;\;\;\;{$\alpha(-1)$}\;\;\;&\;\;\;\;\;{$\beta$}\\  
    \hline
    MoS$_2$&\;\;\;\;\;$4.91$&\;\;\;\;$4.10$&\quad$4.49$\\
    WS$_2$&\;\;\;\;\;$7.73$&\;\;\;\;$4.77$&\quad$6.07$ \\
    MoSe$_2$&\;\;\;\;\;$5.09$&\;\;\;\;$3.99$&\quad$4.51$\\
    WSe$_2$&\;\;\;\;\;$8.27$&\;\;\;\;$4.63$&\quad$6.19$ \\
    \hline
    \hline
\end{tabular}
\end{table}

For the inter-valley exciton exchange interaction,
the L-R part of the exchange
  interaction between the initial exciton state $|m,n,{\bf P}, {\bf K}\rangle$
and the final one $|m',n',{\bf P'}, {\bf K}'\rangle$ is written as 
\begin{equation}
H^{\rm L\mbox{-}R}_{{\bf K}\mbox{-}{\bf K}'}=-C\delta_{\bf P, P'}\frac{P^2_{+}}{|{\bf P}|}\left(\begin{array}{cccc}
\beta & 0 & 0 &\alpha(1)\\
0 & 0 & 0  & 0\\
0 & 0  & 0  & 0\\
\alpha(-1) & 0 & 0 & \beta
\end{array}\right)，
\label{longrange_inter}
\end{equation}
with $P_{\pm}=P_x\pm iP_y$.                                           

From Eqs.~(\ref{longrange_intra}) and (\ref{longrange_inter}), both the intra-
  and inter-valley L-R exchange interactions can cause the valley
  depolarization by the
MSS mechanism.\cite{Sham1,Sham2} In the MSS
mechanism, similar to the DP
mechanism,\cite{DP} the L-R exchange interaction provides a ${\bf P}$-dependent
effective magnetic field ${\bf \Omega}({\bf P})$, around which the ``spins'' of the
exciton with different center-of-mass momentums
process
 with different frequencies, i.e.,
the inhomogeneous broadening.\cite{broadening1,broadening2} This inhomogeneous broadening can cause
a free-induction-decay due to the destructive interference without the exciton
scattering. When there exists exciton scattering with the momentum relaxation
time denoted by $\tau_{P}^*$, the system can be divided into
the weak and strong scattering regimes: in the weak scattering regime with
$\langle|{\bf \Omega}({\bf P})|\rangle\tau_{P}^*\gtrsim 1$, the momentum scattering
opens a spin relaxation channel and the exciton ``spin''
relaxation time $\tau_s\propto \tau_{P}^*$; in the strong scattering regime with
 $\langle|{\bf \Omega}({\bf P})|\rangle\tau_{P}^*\ll 1$, the momentum scattering
 suppresses the inhomogeneous broadening and $\tau^{-1}_s=\langle
 {\bf \Omega}^2({\bf P})\rangle\tau^*_{P}$. Here, $\langle...\rangle$ denotes the ensemble average. 

For the intra-valley exchange interaction, from Eq.~({\ref{longrange_intra}}), when there exists a large splitting of the valence
bands
 the intra-valley
  depolarization channel by the MSS mechanism is nearly forbidden due to the
  detuning effect.\cite{Haug} 
For the inter-valley exciton
  exchange interaction, from
  Eq.~({\ref{longrange_inter}}), there exist only matrix elements between the bright
exciton states. Accordingly, it can cause the valley depolarization due
to the MSS mechanism.  
Specifically, this inter-valley depolarization channel can be efficient
 between two energy-degenerate exciton states when $|{\bf P}|\neq 0$.

\subsubsection{S-R part}
We then express the S-R part of the exchange interaction, which can
exist not only in a single valley, but also between two different valleys. Their matrix
elements between the two exciton states $|m,n,{\bf P}, {\rm {\bf k}_0}\rangle$
and $|m',n',{\bf P}', {\rm {\bf k}'_0}\rangle$ are
 expressed as
\begin{eqnarray}
\nonumber
&&B^{\rm S\mbox{-}R}_{{\rm {\bf k}_0}\mbox{-}{\rm {\bf k}'_0}}=\frac{1}{S}\delta_{{\bf P},{\bf P}'}|\phi_{1s}^{\rm
    2D}(0)|^2\int \big[\Psi^{m'}_{\rm {\bf k}'_0}({\bf r}_1)\big]^*
\big[\Theta \tilde{\Psi}^{n}_{\rm {\bf k}_0}({\bf
  r}_2)\big]^*\\
&&\mbox{}\times U({\bf r}_1-{\bf r}_2)\big[\Theta\tilde{\Psi}^{n'}_{\rm {\bf k}'_0}({\bf
  r}_1)\big]\Psi^m_{\rm {\bf k}_0}({\bf
  r}_2)d{\bf r}_1d{\bf r}_2.
\label{short_range}
\end{eqnarray}
Here, $U({\bf r}_1-{\bf r}_2)={e^2}/\big({4\pi\kappa\varepsilon_0 |{\bf
      r}_1-{\bf r}_2|}\big)$ is the Coulomb potential; $\rm {\bf k}_0({\bf k}'_0)$ represents the
  {\bf K} or {\bf K}$'$.

Accordingly, by using the conduction band and valence band wave
 functions Eq.~(\ref{basis}),
the S-R part of the exchange
  interaction between the initial exciton state $|m,n,{\bf P}, {\bf k}_0\rangle$
and the final one $|m',n',{\bf P'}, {\bf k}'_0\rangle$ is written as
\begin{equation}
H_{{\rm {\bf k}_0}\mbox{-}{\rm {\bf k}'_0}}^{\rm S\mbox{-}R}=\Xi\delta_{\bf P, P'}\left(\begin{array}{cccc}
1 & 0 & 0 & 1\\
0 & 0 & 0 & 0\\
0 & 0 & 0 & 0\\
1 & 0 & 0 & 1
\end{array}\right).
\label{shortrange}
\end{equation}
Here,
\begin{eqnarray}
\nonumber
 \Xi&=&\frac{1}{S}|\phi_{1s}^{\rm 2D}(0)|^2\frac{e^2}{4\pi
    \kappa\varepsilon_0}\frac{1}{2\pi}\int \Big\{|A({\bf q})|^2+|B({\bf
    q})|^2\\
&&\mbox{}+i\big[A^*({\bf q})B({\bf q})-A({\bf q})B^*({\bf q})\big]\Big\}\frac{d{\bf
    q}}{2q},
\label{S_R}
\end{eqnarray}
 in which 
\begin{equation}
A({\bf q})=\langle d_{z^2}|e^{i{\bf q}\cdot{\bf r}}|d_{x^2-y^2}\rangle,~~B({\bf q})=\langle d_{z^2}|e^{i{\bf q}\cdot{\bf r}}|d_{xy}\rangle.
\label{Aq}
\end{equation}  
 
 From Eq.~(\ref{shortrange}), for both the intra- and inter-valley exchange
 interactions, there exist only matrix elements between the bright exciton states, and hence both the
 intra- and inter-valley S-R exchange interactions can only cause
 the bright exciton transition.
 By considering
the large splitting of the valence bands, the intra-valley depolarization channel
due to the intra-valley S-R exchange interaction is nearly forbidden, and
hence only the inter-valley S-R exchange
interaction can contribute to the valley depolarization. 
%\Red{Nevertheless, for
% bilayer MoS$_2$,\cite{absorption_Mak,many_body} the inversion symmetry imposes doubly degenerate conduction and
%  valence bands in both the K and K$'$ valleys, the intra-valley SR
%  exchange interaction can be effective for the exciton dynamics.}  
 
\subsection{Valley depolarization due to the inter-valley e-h exchange interaction} 
\label{Model_B}
\subsubsection{Model and KSBEs}
 From Sec.~{\ref{Model_A}}, we conclude that only the inter-valley e-h
   exchange interaction can cause the valley depolarization efficiently. For the
   A-exciton pumped, the exchange interaction includes the L-R and S-R parts for
   the two energy-degenerate bright
 exciton states
 $\mid \uparrow_e,\downarrow_h,{\bf P}, {\bf
   K}\rangle$ and $\mid \downarrow_e,\uparrow_h,{\bf P}', {\bf K}'\rangle$. By referring to $\mid\uparrow_e,\downarrow_h,{\bf P}, {\bf
   K}\rangle$ and $\mid\downarrow_e,\uparrow_h,{\bf P}', {\bf K}'\rangle$ as
 ``spin''-up $\mid\Uparrow\rangle$ and ``spin''-down $\mid\Downarrow\rangle$
 states, their matrix elements are denoted by $H^{\rm A}_{{\rm
    {\bf K}}\mbox{-}{\rm {\bf K}'}}={\bf \Omega}({\bf P})\cdot {\bf s}$ in the exciton
``spin'' space, with the
effective magnetic field reading 
\begin{equation}
  {\bf \Omega}({\bf P})=\big(-C\alpha(1)\frac{P^2_x-P^2_y}{|{\bf P}|}+\Xi,2C\alpha(1)
  \frac{P_xP_y}{|{\bf P}|},0\big).
\label{eff_mag}
\end{equation}
Obviously, the L-R (S-R) part of the exchange interaction acts as an in-plane ${\bf P}$-dependent (static)
magnetic field.

With the effective magnetic field, the inter-valley A-exciton dynamics
 can be described by the KSBEs:\cite{Sham1,Sham2,broadening1,broadening2}
\begin{equation}
  \partial_t\rho({\bf P},t)=\partial_t\rho({\bf
      P},t)|_{\rm coh}+\partial_t\rho({\bf P},t)|_{\rm  scat}.
\label{ksbe}
\end{equation}
In these equations, $\rho({\bf P},t)$  represent the $2\times2$ density matrices of
A-exciton with center-of-mass momentum ${\bf P}$ at time $t$,
 in which the diagonal elements $\rho_{s,s}({\bf P},t)$ describe the A-exciton distribution
 functions and the off-diagonal elements $\rho_{s,-s}({\bf P},t)$ 
represent the ``spin'' coherence. In the
  collinear space, the coherent term is given by 
%($\hbar\equiv 1$ throughout
%  this paper) 
\begin{equation}
\partial_t\rho({\bf P},t)|_{\rm
   coh}=-\frac{i}{\hbar}\big[{\bf \Omega}({\bf P})\cdot {\bf s},\rho({\bf
  P},t)\big],
\end{equation}
where $[\ ,\ ]$ denotes the commutator.
 The scattering terms
 $\partial_t\rho({\bf P},t)|_{\rm  scat}$ include the inter-exciton scattering,
 exciton-phonon scattering and exciton-impurity scattering. Here, for
 simplicity, we only include the exciton-impurity scattering,\cite{Sham1} which is written as 
 \begin{equation}
\partial_t\rho({\bf P},t)|_{\rm
   scat}=\sum_{{\bf P}'}W_{{\bf P}{\bf P}'}\big[\rho({\bf P}',t)-\rho({\bf
  P},t)\big].
\label{scat}
\end{equation}
Here, $W_{{\bf P}{\bf P}'}$ represents the momentum scattering rate.

By solving the KSBEs, one
obtains the evolution of the valley polarization
$P(t)=\sum_{\bf P}\mbox{Tr}[\rho({\bf P},t) s_z]/n_{ex}$ with $n_{ex}$=$\sum_{\bf
  P}$Tr[${\rho({\bf P},t)}]$ being the density of the A-exciton. 
According to the pump-probe experiment,\cite{many_body,CD} the initial condition is set to be
\begin{equation}
\rho_{s,s}({\bf P},0)=\alpha_s\exp\Big\{-\big[\varepsilon({\bf P})-\varepsilon_{\rm pump}\big]^2/(2\delta^2_{\varepsilon})\Big\}
\end{equation}
and $\rho_{s,-s}({\bf P},0)=0$.
Here, $\varepsilon({\bf P})=\hbar^2{\bf P}^2/(2m^*)$ is the exciton kinetic energy with
$m^*$ being the exciton effective mass; $\varepsilon_{\rm pump}$ is
 the energy of pulse center in reference to the band
minimum and $\delta_{\varepsilon}=\hbar/\delta_{\tau}$ with $\delta_{\tau}$
denoting the pulse width; \begin{equation}
\alpha_s=\frac{n_{\rm pump,s}}{\sum_{\bf P}\exp\Big\{-\big[\varepsilon({\bf P})-\varepsilon_{\rm pump}\big]^2/(2\delta^2_{\varepsilon})\Big\}},
\end{equation}
with $n_{\rm pump,s}$ being the density of A-exciton with ``spin" $s$ after
excitation. In the PL experiment or the pump-probe experiment, according to the chiral optical valley
selection rule, we set $n_{\rm
  pump,\Uparrow}=n_{\rm ex}$ and $n_{\rm
  pump,\Downarrow}=0$. 

\subsubsection{Results}

In this part, we look into the current valley polarization experiments in
monolayer MoS$_2$ with A-exciton pumped: the time-resolved valley
 polarization measurement,\cite{absorption_Marie} the
pump-probe experiment\cite{many_body,CD} and the steady-state
 PL polarization measurement.\cite{direct_gap1,valley_wang20,absorption_Mak,valley_wang24,
absorption_Kioseoglou,valley_wang18,absorption_wang15} Their
theoretical explanations are summarized below based on
the KSBEs
[Eq.~(\ref{ksbe})]. The material parameters in our computation are listed in
Table~\ref{material}.
\begin{table}[htb]
  \caption{Material parameters used in the computation.}
  \label{material} 
  \begin{tabular}{l l l l}
    \hline
    \hline
    $\kappa$&\;\;\;\;\;$3.43^a$&\;\;\;\;$a_B$(nm)&\quad$2.0^{c}$\\
    $m^*/m_0$&\;\;\;\;\;$0.28^{b}$&\;\;\;\;$\tau_P^*$ (fs)&\quad$6.0$ \\
    $n_{\rm ex}$ (cm$^{-2}$)&\;\quad$10^{10}$&\;\;\;\;$\alpha(1)$ (${\rm \AA}^2$)&\quad$4.91$\\
    \hline
    \hline
\end{tabular}\\
 $^a$ Ref.~\onlinecite{directgap_wang23}. \quad$^b$
 Refs.~\onlinecite{splitting_wang39,splitting_wang22}.
 \quad $^c$ Refs.~\onlinecite{2D,radius1,radius2,small}. 
\end{table}

We first study the initial evolution of the valley
polarization in the time-resolved polarization
measurement in monolayer MoS$_2$ carried out by Lagarde
{\it et al.},\cite{absorption_Marie} in which the
emergence of the A-exciton in the K$'$ valley is almost immediate with the A-exciton pumped in 
the K valley. In the experiment, the pulse-center energy is away from the
A-exciton resonance energy by $\varepsilon_{\rm
  pump}\approx 100$~meV and the laser pulse width $\delta_t\approx 1.6$
ps.\cite{absorption_Marie} With this pulse, the center-of-mass
momentum $|{\bf P}|=\sqrt{2m^*\varepsilon_{\rm
  pump}}/\hbar$ of the A-exciton is large. From Eq.~(\ref{eff_mag}), the precession frequency due to the L-R
exchange interaction between the two
exciton ``spin'' states is estimated to be
\begin{equation}
{\bf \omega}({\bf P})\approx \sqrt{5}C\alpha(1)|{\bf P}|/\hbar,
\end{equation}
which is proportional to $|{\bf P}|$. Obviously, when $|{\bf P}|$ is large (small), the
precession frequency between the two exciton ``spin'' states is large (small) and the L-R
exchange interaction causes {\em fast} ({\em slow}) inter-valley exciton
precession. Specifically, when $|{\bf P}|=0$, the inter-exciton precession due
to the L-R exchange interaction is forbidden and the inter-valley exciton
precession time is expected to be very long. Accordingly, in the experiment of Lagarde
{\it et al.},\cite{absorption_Marie}  due to the large initial A-exciton center-of-mass momentum, 
the inter-valley exciton precession time is estimated to be $T=\pi/{\bf
  \omega}({\bf P})=13$ fs, which is much shorter than the uncertainty of time
origin, i.e., 700 fs in the experiment.\cite{absorption_Marie} Therefore, the observation of the
excitonic signal in the K$'$ valley is immediate after the A-exciton pumped in K
valley in the
experiment.\cite{absorption_Marie} Here, the S-R exchange interaction is not
considered. Only the
L-R exchange interaction can well explain the experiment.\cite{absorption_Marie}
For the S-R exchange interaction, unlike the L-R component, so far there lacks the
material parameter $\Xi$ [Eq.~(\ref{S_R})].
Furthermore, according to the experience in
semiconductors, the S-R exchange interaction is much
smaller than the L-R one.\cite{broadening2,dot,GaAs} Therefore, the S-R exchange interaction is speculated
to be negligible here.

\begin{figure}[htb]
  {\includegraphics[width=8cm]{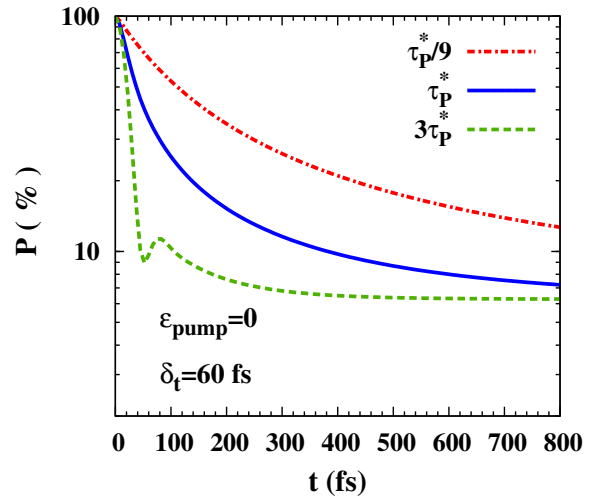}}
  \caption{(Color online) Evolution of the valley polarization when A-exciton is
    resonantly pumped in the K valley with different momentum relaxation
    times $\tau^*_P/9$, $\tau^*_P$ and $3\tau^*_P$, respectively.
 The pulse width is chosen to be $\delta_t=60$ fs according to Ref.~\onlinecite{many_body}.}
\label{figyw2}
\end{figure}

We then investigate the dynamics of the valley
polarization in the pump-probe experiments for monolayer MoS$_2$ based on the
KSBEs with the A-exciton resonantly pumped in the K valley.\cite{many_body,CD} In our
calculation,
 the momentum relaxation time $\tau^*_{P}$ in Table~\ref{material} is obtained based on the elastic
scattering approximation as a first step in the investigation,\cite{Sham1} which
can be varied by tuning $W_{{\bf P}{\bf P}'}$ [Eq.~(\ref{scat})]
in the calculation. Its value is estimated to be 6 fs by considering the measured broadening of the A exciton
 energy 
 $\Gamma\approx 110$~meV at 4~K with $\tau^*_P\approx
 \hbar/\Gamma$.\cite{absorption_Marie,Glazov} By setting $\varepsilon_{\rm
   pump}=0$ eV and $\delta_t=60$ fs according to Mai {\it et al},\cite{many_body} with the material parameters in
 Table~\ref{material},
 the evolution of the valley polarization with different scattering strengths can be
obtained by numerically solving the KSBEs, shown in
Fig.~\ref{figyw2}. 

From Fig.~\ref{figyw2}, several features of the dynamics
of the valley polarization can be obtained, which are in good agreement with the
experimental observations
    in the pump-probe set-up.\cite{many_body,CD} It is
observed that when the A-exciton is resonantly pumped in the K valley, even
at the time overlap of the pump and probe pulses, there are excitonic signals of
the A-exciton in the K$'$ valley. Here, in our computation, it is shown that with the A-exciton
    resonantly pumped in the K valley, when the momentum scattering is
    relatively weak (the green dashed curve with momentum relaxation time
    $3\tau^*_P$), it takes only several femtoseconds
 for the transition of the A-exciton from the K valley to the K$'$ valley; even the momentum scattering is
    relatively strong, this inter-valley transition time for the A-exciton is
    still in
    the order of tens of femtoseconds. This is also due to the large center-of-mass
    momentum of the exciton, the effective
    magnetic field due to the L-R exchange
    interaction can cause fast inter-exciton ``spin'' precession. Furthermore, it is also reported in the
    experiment\cite{many_body} that the polarization anisotropy in
the A transition is completely lost in about 400 fs, which indicates the valley
depolarization time is hundreds of femtoseconds. Our calculation also
shows that no matter the momentum scattering is
relative weak (the green dashed curve with momentum relaxation time
$3\tau^*_P$) or strong (the red chain curve
with momentum relaxation time $\tau^*_P/9$),
 the valley depolarization times are in the order of hundreds of
 femtoseconds. Moreover, in the experiment,\cite{many_body} it shows that the
total valley polarization does not completely relax for about 10 ps, which
hints the existence of the residue valley polarization which lasts for a very long time. In
our computation, it is also observed that there are residue valley polarizations (about $10\%$) lasting for
 several picoseconds. This residue valley polarization originates from the
 small effective magnetic field due to the L-R exchange interaction with
   small $|\bf P|$ and hence long valley depolarization time, in contrast to the fast
 inter-exciton precession with large $|{\bf P}|$.
 Furthermore, due to the energy relaxation of the excitons, with
 the increase of the ratio of the excitons with small $|{\bf P}|$, the residue
 valley polarization increases.

Finally, we address the series of steady-state
measurements of the PL polarization.\cite{absorption_Mak,valley_wang24,
absorption_Kioseoglou,valley_wang18,absorption_wang15} In these
experiments, a wide range of residue valley polarizations from $30\%$ to $100\%$
are reported with the A-exciton resonantly pumped at low
temperature.\cite{absorption_Mak,valley_wang24,
absorption_Kioseoglou,valley_wang18,absorption_wang15} These
steady-state residue PL polarizations can be estimated by the rate
equations,\cite{absorption_Mak,absorption_Kioseoglou,valley_wang18,absorption_Marie,Yu3}
a simplified KSBEs, with the valley depolarization time and
exciton lifetime known. We point out that it has been addressed by us
with the S-R exchange interaction only\cite{Yu3} and by 
Glazov {\it et al.} very recently with the L-R exchange
interaction.\cite{Glazov} However, based on the above understanding, 
we conclude that the L-R exchange interaction should
be more important than the S-R one in the valley depolarization.

\section{Conclusion and discussion}
\label{summary}
In conclusion, we have investigated the valley depolarization due to the e-h
exchange interaction in monolayer
MoS$_{2}$. Both the L-R and S-R parts of the exchange interactions for
the inter- and intra-valley e-h interactions are calculated. We
find that both the L-R and S-R exchange interactions can cause the
inter- and intra-valley bright exciton transitions.
 However, the intra-valley bright exciton transition channel
is nearly forbidden due to the large splitting of the valence
bands. For the inter-valley bright exciton transition, we show that for the 
A-exciton with large center-of-mass momentum, the L-R exchange 
interaction can cause the {\em fast} inter-valley exciton transition.
 This explains the fast emergence of the excitonic signal in the K$'$
valley with the A-exciton pumped in K valley in the experiments.\cite{absorption_Marie,many_body,CD}
However, for the 
A-exciton with small center-of-mass momentum, the inter-valley exciton
transition is relatively {\em slow} and this leads to the existence of the residue valley
polarization which lasts for tens of picoseconds in the experiments.\cite{many_body,CD}
As for the S-R exchange interaction, whose strength is unavailable due to
  lack of the material parameter in the literature, it is speculated to be
  negligible.
 More investigations are needed to further
clarify this problem.

Finally, we address other possible valley depolarization mechanisms in the
  literature. For the DP and EY
  mechanisms, only the extrinsic influences can cause the relaxation of
  the out-of-plane component of the electron/hole spin
  but with low efficiency, which cannot cause the inter-valley exciton
  transition effectively.  Apart from the inter-valley exciton transition,
  there also exist
  other arguments for the possible cause of the valley
  depolarizations in the experiments.  It was argued that the out-of-plane
  component of the hole spin can
  relax after scattering to the $\Gamma$ valley through the DP or EY mechanisms.\cite{many_body} However, this
  is impossible as $\hat{s}_z$ is also a good quantum number for the $\Gamma$
  valley, apart from the fact that there is no effective relaxation channel
 for the electron spin.\cite{splitting_wang39,tightbinding}
 In addition, tight-binding simulations show that the 
  disordered defects can weaken the chiral optical valley
selection rule when the excitation is away from the vicinity of the high
symmetry K(K$'$) point.\cite{CD} However, only very strong disorder can significantly
decrease the valley polarization,\cite{CD} which is unlikely in the clean
samples.

\begin{acknowledgments}
This work was supported
 by the National Natural Science Foundation of China under Grant
No. 11334014, the National Basic Research Program of China under Grant No.
2012CB922002 and the Strategic Priority Research Program 
of the Chinese Academy of Sciences under Grant
No. XDB01000000. One of the
authors (TY) would like to thank
L. Wang for valuable discussions. 
\end{acknowledgments}

 \begin{appendix}
 \section{Exciton Hamiltonian}
 \label{AA}
 In this appendix, based on the lowest four band ${\bf
   k}\cdot {\bf p}$ Hamiltonian [Eq.~(\ref{hamiltonian})], we give the explicit
 form of the 
 exciton Hamiltonian $H^{\rm eh}_{m'n'\atop mn}{{\bf r}_1'~{\bf r}_2'\choose {\bf r}_1~{\bf
    r}_2}$ for the direct exciton envelop function $F_{mn}({\bf r}_1,{\bf
    r}_2)$ in the coordinate space, where $m (m')$ and $n (n')$ represent the
  band indexes in the K or K$'$ valley labeled by the
  electron spin for the conduction band and hole spin for the valence band,
  respectively.\cite{exchange,Sham1,dot}
 The eigenequation
  expressed by the exciton Hamiltonian for the exciton envelop function
  satisfies
\begin{equation}
\sum_{mn}\int d{\bf r}_1d{\bf r}_2H^{\rm eh}_{m'n'\atop mn}{{\bf r}_1'~{\bf r}_2'\choose {\bf r}_1~{\bf
    r}_2} F_{mn}({\bf r}_1,{\bf
    r}_2) =E F_{m'n'}({\bf r}_1',{\bf
    r}_2'),
\end{equation}
where
\begin{eqnarray}
\nonumber
&&H^{\rm eh}_{m'n'\atop mn}{{\bf r}_1'~{\bf r}_2'\choose {\bf r}_1~{\bf
    r}_2}=\big[H^e_{m'm}({\bf k}_1)\delta_{n'n}+H^h_{n'n}({\bf
  k}_2)\delta_{m'm}\\
\nonumber
&&\mbox{}+U^{\rm eh}({\bf r}_1-{\bf r}_2)\delta_{m'm}\delta_{n'n}\big]\delta({\bf
  r}_1-{\bf r}_1')\delta({\bf
  r}_2-{\bf r}_2')\\
&&\mbox{}+H^{\rm ex}_{m'n'\atop mn}{{\bf r}_1'~{\bf r}_2'\choose {\bf r}_1~{\bf
    r}_2}.
\label{effective_H}
\end{eqnarray}
Here, ${\bf k}=-i\nabla$, 
\begin{equation}
U^{\rm eh}({\bf r}_1-{\bf r}_2)=-\frac{e^2}{4\pi \varepsilon_0\kappa|{\bf
    r}_1-{\bf r}_2|},
\end{equation}
\begin{eqnarray}
\nonumber
&&H^e_{m'm}({\bf k}_1)\\
\nonumber
&&=E_m({\bf k}_0)+\frac{\hbar^2}{2m_0^2} \sum_{m''} \big[{\bf
  k}_1\cdot {\bgreek \pi}_{m'm''}({\bf k}_0)\big]\big[{\bf k}_1\cdot {\bgreek
  \pi}_{m''m}({\bf k}_0)\big]\\
&&\mbox{}\times\big[\frac{1}{E_{m}({\bf k}_0)-E_{m''}({\bf k}_0)}+\frac{1}{E_{m'}({\bf k}_0)-E_{m''}({\bf k}_0)}\big],
\label{effective-mass}
\end{eqnarray}
and
\begin{equation}
H^h_{n'n}({\bf k}_2)=-H^e_{\Theta n \Theta n'}(-{\bf k}_2),
\end{equation}
We have ${\bgreek \pi}={\bf p}+\frac{\hbar}{4m_0^2c^2}[{\bgreek
  \sigma}\times (\nabla V_0)]$ with $V_0$ denoting the lattice
potential. ${\bgreek \pi}_{ss'}({\bf k}_0)$ stands for the matrix elements of $\bgreek \pi$
between two Bloch wavefunctions in the $s$ and $s'$ bands ($s$ and $s'$ are electron
spins). The nonzero expressions of ${\bgreek \pi}_{ss'}({\bf k}_0)$ in the K or K$'$
valley can be obtained from the Hamiltonian Eq.~({\ref{hamiltonian}}).
For the K ($\tau=1$) or K$'$ ($\tau=-1$) valley,
\begin{equation}
\langle \uparrow_c|\pi_x|\downarrow_h\rangle=\langle
\downarrow_c|\pi_x|\uparrow_h\rangle=\tau m_0at/\hbar
\end{equation}
and
\begin{equation}
\langle \uparrow_c|\pi_y|\downarrow_h\rangle=\langle
\downarrow_c|\pi_y|\uparrow_h\rangle=-im_0at/\hbar.
\end{equation}

The electron-hole exchange interaction Hamiltonian is divided into L-R and S-R
parts:
\begin{equation}
H^{\rm ex}_{m'n'\atop mn}{{\bf r}_1'~{\bf r}_2'\choose {\bf r}_1~{\bf
    r}_2}=H^{\rm LR}_{m'n'\atop mn}{{\bf r}_1'~{\bf r}_2'\choose {\bf r}_1~{\bf
    r}_2}+H^{\rm SR}_{m'n'\atop mn}{{\bf r}_1'~{\bf r}_2'\choose {\bf r}_1~{\bf
    r}_2}.
\end{equation} 
For the L-R exchange interaction,
\begin{eqnarray}
\nonumber
&&H^{\rm LR}_{m'n'\atop mn}{{\bf r}_1'~{\bf r}_2'\choose {\bf r}_1~{\bf
    r}_2}\\
\nonumber
&&=-\sum_{\alpha\beta}Q^{{\bf k}_0\mbox{-}{\bf k}_0'}_{m' \Theta n \atop \Theta n'
  m}\Big{|}_{\alpha\beta}\frac{\partial^2}{\partial {\bf r}_1^{\alpha}\partial {\bf
    r}_1^{\beta}}U({\bf r}_1-{\bf r}_2')\delta({\bf r}_1-{\bf
  r}_2)\delta({\bf r}_1'-{\bf r}_2'),\\
\end{eqnarray}
with 
\begin{eqnarray}
&&Q^{{\bf k}_0\mbox{-}{\bf k}_0'}_{m' \Theta n \atop \Theta n'
  m}\Big{|}_{\alpha\beta}=\frac{\hbar^2}{m_0^2}\pi_{m'\Theta n'}^{\alpha}({\bf k}_0)\pi_{\Theta n
    m}^{\beta}({\bf k}'_0)\\
\nonumber
&&\mbox{}\times\frac{1}{\big[E_m({\bf k}_0)-E_n({\bf k}_0)\big]\big[E_{m'}({\bf k}'_0)-E_{n'}({\bf k}'_0)\big]}
\end{eqnarray}
and $\alpha$ ($\beta$) denoting $x$ or $y$.
For the S-R exchange interaction, 
\begin{equation}
H^{\rm SR}_{m'n'\atop mn}{{\bf r}_1'~{\bf r}_2'\choose {\bf r}_1~{\bf
    r}_2}=SU_{m' \Theta n \atop \Theta n'
  m}\delta({\bf r}_1-{\bf r}_2)\delta({\bf r}_1-{\bf r}_1')\delta({\bf r}_2-{\bf
r}_2'),
\end{equation}
with 
\begin{eqnarray}
\nonumber
&&U_{m' \Theta n \atop \Theta n'm}=\frac{1}{S^2}\int d{\bf r}_1d{\bf r}_2
\big[\Psi^{m'}_{{\bf k}_0'}({\bf r}_1)\big]^*\big[\Theta\tilde{\Psi}^n_{{\bf k}_0}({\bf
  r}_2)\big]^*\\
&&\mbox{}\times U({\bf r}_1-{\bf r}_2)\big[\Theta \tilde{\Psi}^{n'}_{{\bf k}_0'}({\bf
  r}_1)\big]\Psi^m_{{\bf k}_0}({\bf r}_2).
\end{eqnarray}
\end{appendix}


\begin{thebibliography}{0}

%electrical property
\bibitem{ele_experiment} B. Radisavljevic, A. Radenovic, J. Brivio,
  V. Giacometti, and A. Kis, Nature Nanotech. {\bf 6}, 147 (2011).
\bibitem{ele_theory} Y. Yoon, K. Ganapathi, and S. Salahuddin, Nano Lett. {\bf 11},
3768 (2011).
\bibitem{electric_3} H. Liu and P. D. Ye, IEEE Electron Dev. Lett. {\bf 33}, 546
  (2012).
%valley_polarization
\bibitem{valley_wang6} W. Yao, D. Xiao, and Q. Niu, Phys. Rev. B {\bf 77}, 235406
(2008).
\bibitem{valley_wang18} T. Cao, G. Wang, W. Han, H. Ye, C. Zhu, J. Shi, Q. Niu,
P. Tan, E. Wang, B. Liu, and J. Feng, Nature Commun. {\bf 3}, 887 (2012).
\bibitem{valley_wang20} D. Xiao, G. B. Liu, W. Feng, X. Xu, and W. Yao, Phys.
Rev. Lett. {\bf 108}, 196802 (2012).
\bibitem{valley_wang24} G. Sallen, L. Bouet, X. Marie, G. Wang, C. R. Zhu, W.
P. Han, Y. Lu, P. H. Tan, T. Amand, B. L. Liu, and B. Urbaszek, Phys. Rev. B {\bf 86}, 081301(R) (2012).
%Sallen 90%

%gap
\bibitem{direct_gap1} A. Splendiani, L. Sun, Y. Zhang, T. Li, J. Kim, C. Y.
Chim, G. Galli, and F. Wang, Nano Lett. {\bf 10}, 1271 (2010).
\bibitem{directgap_wang11} Z. Y. Zhu, Y. C. Cheng, and U. Schwingenschl\"ogl, Phys. Rev. B {\bf 84}, 153402 (2011).
\bibitem{direct_gap3} K. Kaasbjerg, K. S. Thygesen, and K. W. Jacobsen, Phys.
Rev. B {\bf 85}, 115317 (2012).
\bibitem{directgap_wang23} T. Cheiwchanchamnangij and W. R. L. Lambrecht, Phys.
Rev. B {\bf 85}, 205302 (2012).
\bibitem{direct_gap5} H. Shi, H. Pan, Y. W. Zhang, and B. I. Yakobson, Phys.
Rev. B {\bf 87}, 155304 (2013).
\bibitem{direct_gap6} X. Li, J. T. Mullen, Z. Jin, K. M. Borysenko, M. B.
Nardelli, and K. W. Kim, Phys. Rev. B {\bf 87}, 115418 (2013).

%valence-band splitting
\bibitem{splitting_wang27} E. S. Kadantsev and P. Hawrylak, Solid State Commun.
{\bf 152}, 909 (2012).
\bibitem{splitting_wang31} K. Ko\'smider and J. F. Rossier, Phys. Rev. B {\bf 87}, 075451
(2013).
\bibitem{splitting_wang37} H. Ochoa and R. Rold\'an, Phys. Rev. B {\bf 87}, 245421 (2013).
\bibitem{splitting_wang38} F. Zahid, L. Liu, Y. Zhu, J. Wang, and H. Guo, AIP
  Advances {\bf 3}, 052111 (2013).

%absorption
\bibitem{absorption_Mak} K. F. Mak, K. He, J. Sahn, and T. F. Heinz, Nature
  Nanotech. {\bf 7}, 494 (2012).
\bibitem{absorption_Kioseoglou} G. Kioseoglou, A. T. Hanbicki, M. Currie, A. L. Friedman, D. Gunlycke,
and B. T. Jonker, Appl. Phys. Lett. {\bf 101}, 221907 (2012).
\bibitem{absorption_wang15} H. Zeng, J. Dai, W. Yao, D. Xiao, and X. Cui, Nature
Nanotech. {\bf 7}, 490 (2012).
\bibitem{absorption_Marie} D. Lagarde, L. Bouet, X. Marie, C. R. Zhu, B. L. Liu,
  T. Amand, and B. Urbaszek, Phys.
Rev. Lett. {\bf 112}, 047401 (2014).

%pump-probe
\bibitem{many_body} C. Mai, A. Barrette, Y. Yu, Y. G. Semenov, K. W. Kim,
  L. Cao, and K. Gundogdu, Nano Lett. {\bf 14}, 202 (2014).
\bibitem{CD} Q. Wang, S. Ge, X. Li, J. Qiu, Y. Ji, J. Feng, and D. Sun, ACS Nano
  {\bf 7}, 11087 (2013).

%spin relaxation
\bibitem{DP} M. I. D'yakonov and V. I. Perel', Zh. Eksp. Teor. Fiz. {\bf 60}, 1954
  (1971) [Sov. Phys. JETP {\bf 33}, 1053 (1971)].
\bibitem{Yafet} Y. Yafet, Phys. Rev. {\bf 85}, 478 (1952).
\bibitem{Elliott} R. J. Elliott, Phys. Rev. {\bf 96}, 266 (1954).

\bibitem{wang_relaxation1} L. Wang and M. W. Wu, Phys. Lett. A, 2014, in
  press; arXiv:1305.3361v2.
\bibitem{splitting_wang39} A. Korm\'anyos, V. Z\'olyomi, N. D. Drummond, P. Rakyta,
G. Burkard, and V. I. Fal'ko, Phys. Rev. B {\bf 88}, 045416
(2013).
\bibitem{tightbinding} H. Rostami, A. G. Moghaddam, and R. Asgari, Phys. Rev. B
  {\bf 88}, 085440 (2013). 
\bibitem{referee} H. Ochoa and R. Rold\'an, Phys. Rev. B {\bf 87}, 245421 (2013).
\bibitem{flexural} H. Ochoa, F. Guinea, and V. I. Fal'ko, Phys. Rev. B
  {\bf 88}, 195417 (2013). 
\bibitem{wang_relaxation2} L. Wang and M. W. Wu, Phys. Rev. B {\bf 89}, 115302 (2014). 
%BAP
\bibitem{CD14} T. Korn, S. Heydrich, M. Hirmer, J. Schmutzler, and C. Sch\"uller,
  Appl. Phys. Lett. {\bf 99}, 102109 (2011).

%MSS
\bibitem{Sham1} M. Z. Maialle, E. A. de Andrada e Silva, and L. J. Sham,
  Phys. Rev. B {\bf 47}, 15776 (1993). 
\bibitem{Sham2} A. Vinattieri, Jagdeep Shah, T. C. Damen, D. S. Kim,
  L. N. Pfeiffer, M. Z. Maialle, and L. J. Sham, Phys. Rev. B {\bf 50}, 10868 (1994). 

\bibitem{broadening1} M. W. Wu and C. Z. Ning, Eur. Phys. J. B. {\bf 18}, 373 (2000);
  M. W. Wu, J. Phys. Soc. Jpn. {\bf 70}, 2195 (2001). 
\bibitem{broadening2} M. W. Wu, J. H. Jiang, and M. Q. Weng, Phys. Rep. {\bf 493},
  61 (2010).

\bibitem{exchange} G. E. Pikus and G. L. Bir, Zh. Eksp. Teor. Fiz. {\bf 60}, 195 (1971) [Sov.
Phys. JETP {\bf 33}, 108 (1973)].
\bibitem{dot} H. Tong and M. W. Wu, Phys. Rev. B {\bf 83}, 235323 (2011).

\bibitem{2D} A. Ramasubramaniam,
  Phys. Rev. B {\bf 86}, 115409 (2012).


%detuning
\bibitem{Haug} H. Haug and A.-P. Jauho, {\em Quantum Kinetics in Transport and 
Optics of Semiconductors} (Springer-Verlag, Berlin, 1996).

\bibitem{splitting_wang22} K. Kaasbjerg, K. S. Thygesen, and K. W. Jacobsen, Phys. Rev. B {\bf 85}, 115317
  (2012).
%excitonic
\bibitem{radius1} T. Cheiwchanchamnangij and W. R. L. Lambrecht, Phys.
Rev. B {\bf 85}, 205302 (2012).
\bibitem{radius2} F. J. Crowne, M. Amani, A. G. Birdwell, M. L. Chin,
T. P. O'Regan, S. Najmaei, Z. Liu, P. M. Ajayan, J. Lou,
and M. Dubey, Phys. Rev. B {\bf 88}, 235302 (2013).
 \bibitem{small} C. Zhang, H. Wang, W. Chan, C. Manolaton, and F. Rana,
   arXiv:1402.0263.


\bibitem{GaAs} {\em Numerical Data and Functional Relationships in Science and
Technology}, Landolt-B\"ornstein, New Series, Group III, Vol. 17,
Pt. A, edited by O. Madelung, M. Schultz, and H. Weiss (Springer-
Verlag, Berlin, 1982).

%transitions
\bibitem{Yu3} T. Yu and M. W. Wu, arXiv:1401.0047v2. 
\bibitem{Glazov} M. M. Glazov, T. Amand, X. Marie, D. Lagarde, L. Bouet, and
  B. Urbaszek, arXiv:1403.0108.



\end{thebibliography}
\end{document}